\documentclass[fleqn,twoside]{article}
\usepackage{espcrc2}

\renewcommand{\>}{\rangle} 
\newcommand{\beq}{\begin{equation}}
\newcommand{\eeq}{\end{equation}}
\newcommand{\beqn}{\begin{eqnarray}}
\newcommand{\eeqn}{\end{eqnarray}}

\newcommand{\AmS}{{\protect\the\textfont2
A\kern-.1667em\lower.5ex\hbox{M}\kern-.125emS}}

\title{ 
\vspace{-0.5cm}
\hspace{13.0cm} {\small Bicocca FT-03-29} \\
\vspace{0.2cm}
Chirally improving Wilson fermions\thanks{Partially supported 
by the European Community's Human Potential Programme under contract 
HPRN-CT-2000-00145, Hadrons/Lattice QCD}\thanks{Talk presented by G.C. Rossi}}

\author{R. Frezzotti\address{INFN, Sezione di Milano and 
Dipartimento di Fisica, Universit\`a di Milano ``{\it Bicocca}''\\
Piazza della Scienza 3 - 20126 Milano (Italy)}%
and G.C. Rossi\address{Dipartimento 
di Fisica, Universit\`a di  Roma ``{\it Tor Vergata}'' and INFN, 
Sezione di Roma 2\\Via della Ricerca Scientifica - 00133 Roma (Italy)}}

\begin{document}

\begin{abstract}
It is possible to improve the chiral behaviour and the approach 
to the continuum limit of correlation functions in lattice QCD with 
standard and twisted Wilson fermions by taking arithmetic averages 
of correlators computed in theories regularized with Wilson terms of opposite 
sign. To avoid the problem of the spurious zero-modes of the Wilson--Dirac 
operator, twisted-mass lattice QCD should be used for the actual computation 
of the correlators taking part in the averages. A particularly useful choice 
for the twisting angle is $\pm\pi/2$ where many physical quantities 
(e.g. masses and zero-momentum matrix elements) are automatically improved 
with no need of averaging.
\vspace{1pc}
\end{abstract}

\maketitle
\vspace{-0.2cm}

\section{INTRODUCTION}

In this talk I wish to present a simple strategy~\cite{FR} which allows 
to get from simulations employing standard or twisted Wilson 
fermions lattice data that are free of O($a$) discretization errors 
and have a somewhat smoother and more chiral behaviour near the continuum 
limit than data obtained in unimproved simulations.

Indeed it can be shown that O($a$) discretization effects are absent 
from the average of correlators (Wilson average, $WA$) computed with lattice 
actions having Wilson terms of opposite sign and a common value of the 
subtracted lattice quark mass $m_q=M_0-M_{\rm{cr}}$ ($M_0$ and $M_{\rm{cr}}$ 
are bare and critical quark masses). Thus fully O($a$) improved lattice data 
for energy levels (hence hadronic masses), matrix elements and 
renormalization constants can be obtained, without the need of computing 
anyone of the usual lattice improvement coefficients.

Absence of O($a$) discretization errors in $WA$'s is proved by referring to 
the Symanzik expansion (SE) of connected, on-shell lattice correlators in 
terms of continuum Green functions and exploiting the relations derived 
by matching the ``${\cal{R}}_5$-parity'' of lattice correlators under
\beq
{\cal{R}}_5 :\psi\rightarrow\psi'=\gamma_5 \psi \, ,\quad
\bar{\psi}\rightarrow\bar{\psi}'=-\bar{\psi}\gamma_5 
\label{PSIBAR}\eeq
to the ${\cal{R}}_5$-parities of the related continuum Green functions. 
The latter correspond to the parity of continuum Green functions under
the change of sign of the continuum quark mass parameter. 

Standard Wilson fermions have the well known problem of being afflicted 
by the existence of ``exceptional configurations''. The problem is even 
more worrysome in the present approach because one will always have to face  
the situation in which the relative sign of the coefficients of Wilson 
and quark mass terms are such that they give opposite contributions to the 
real part of the eigenvalues of the Wilson--Dirac (WD) operator. 

The remedy to this situation is to use tm-LQCD~\cite{FGSW}. No one of 
the previous cancellations and improvements is lost by taking averages 
of correlators evaluated with tm-LQCD actions having (twisted) Wilson terms 
of opposite sign.

Moreover, if the special choice $\pm \pi/2$ for the twisting angle 
is made, many interesting physical quantities (e.g.~masses and zero-momentum 
matrix elements) can be extracted from lattice data that are automatically 
O($a$) improved with no need of any $WA$.

\section{THE GENERAL ARGUMENT}

For illustrative purpose we present the key argument in the simple case 
of standard Wilson fermions. We are interested in computing 
($x\equiv \{x_1\neq x_2\neq\ldots\neq x_n\}$)
\beq \langle O(x) \rangle_{(r,M_0)}=
\frac{1}{Z_{\rm{QCD}}^{\rm{L}}} \int {\cal {D}} \mu
\, e^{-[S_{\rm{YM}}^{\rm{L}}+S_{\rm{WF}}^{\rm{L}}]}\, O(x)
\label{EXPVAL}\eeq 
with $O$ a gauge invariant, multi-local and multiplicatively renormalizable 
(m.r.) operator. The subscript $(r,M_0)$ means that the correlator is taken 
with the specified values of the Wilson parameter, $r$, and bare quark mass, 
$M_0$. 

The key observation of this work is that under~(\ref{PSIBAR}) the fermionic
action $S_{\rm{WF}}^{\rm{L}}$ goes into itself, if at the same time we change 
sign to the Wilson term 
(i.e.~to $r$) and $M_0$. In the spirit of spurion analysis, a quick 
way of studying the situation is to momentarily treat $r$ and $M_0$ as 
fictitious fields and consider the combined transformation 
\beq
{\cal{R}}_5^{\rm{sp}}\equiv{\cal{R}}_5\times [r\rightarrow -r]\times 
[M_0\rightarrow -M_0]
\label{SPURM0}\eeq
as a symmetry of the lattice theory. Thus any (m.r.) operator will be either 
even or odd under ${\cal{R}}_5^{\rm{sp}}$, because 
$[{\cal{R}}_5]^2=[{\cal{R}}_5^{\rm{sp}}]^2=1\!\!1$. From this argument one
deduces the relation
\beq 
\langle O(x)\rangle_{(r,M_0)}= (-1)^{P_{{\cal{R}}_5}[O]}\langle
O(x)\rangle_{(-r,-M_0)} \, , \label{KEYRELM0} \eeq 
where $(-1)^{ P_{{\cal{R}}_5}[O] }$ is the ${\cal{R}}_5$-parity of the bare
counterpart of $O$. Eq.~(\ref{KEYRELM0}) follows by performing 
in the functional integral~(\ref{EXPVAL}) the change of fermionic integration 
variables induced by~(\ref{PSIBAR}). 

Already from this simple remark one concludes that 1) the renormalization 
constant of any m.r.\ operator $O$, $Z_O$, must be even under $r\rightarrow -r$
in order for the renormalized operator, $\hat O= Z_O O$, to have the same 
${\cal{R}}_5$-parity as $O$; 2) $M_{\rm{cr}}(r)$ must instead be odd.

As $m_q=M_0-M_{\rm{cr}}$ changes sign under 
$[M_0 \rightarrow - M_0] \times [r \rightarrow -r]$, ${\cal R}_5^{\rm sp}$ 
can also be written as 
\beq
{\cal{R}}_5^{\rm{sp}} = {\cal{R}}_5\times [r\rightarrow -r]\times
[m_q\rightarrow -m_q] \, .
\label{SPUR}\eeq
Correspondingly, the relation~(\ref{KEYRELM0}), which expresses the
implications of the spurionic symmetry ${\cal R}_5^{\rm sp}$ on lattice
correlators, takes the form
\beq
\langle O(x)\rangle_{(r,m_q)}= (-1)^{P_{{\cal{R}}_5}[O]}\langle
O(x)\rangle_{(-r,-m_q)} \, . \label{KEYREL} \eeq
We have now indicated the parameters that specify the fermionic 
action by using, besides $r$, $m_q$ instead of $M_0$. 

In order to discuss the issue of O($a$) improvement we need to make 
reference to the SE of lattice correlators in terms of correlators of the 
continuum theory~\cite{LSSW}. Schematically up to O($a$) terms, one gets 
for the lattice expectation value of a m.r.\ operator ($n_{\ell}>0$)
\beqn
\hspace{-.5cm}&&\langle O(x) \rangle_{(r,m_q)} = [\zeta^{O}_{O}(r)+
am_q\xi^{O}_{O}(r)]\langle O(x) \rangle^{\rm{cont}}_{(m_q)}+
\nonumber\\
\hspace{-.5cm}&&+ a\,\sum_{\ell}(m_q)^{n_{\ell}}\eta^{O}_{O_{\ell}}(r)
\langle O_{\ell}(x)\rangle^{\rm{cont}}_{(m_q)}+{\rm{O}}(a^2)\, .
\label{SCHSYM}\eeqn
It is important to stress that, in order for the formal counting 
of powers of $a$ yielded by the SE to be meaningful, it is necessary 
that one is dealing with expectation values of m.r.\ lattice operators.

At this point one can prove that the Symanzik coefficients $\zeta$ are even 
functions of $r$, while all the others ($\xi$ and $\eta$) are odd. This is 
the consequence of eqs.~(\ref{KEYREL}), (\ref{SCHSYM}), and the 
transformation properties of 
lattice and continuum correlators under the functional change of variables
induced by the transformation ${\cal{R}}_5\times {\cal{D}}_d$, where
\begin{equation}   
{\cal{D}}_d : \left \{\begin{array}{lll}    
U_\mu(x)&\rightarrow U_\mu^\dagger(-x-a\hat\mu) \\
\psi(x)&\rightarrow e^{3i\pi/2}\, \psi(-x)  \\   
\bar{\psi}(x)&\rightarrow e^{3i\pi/2} \, \bar{\psi}(-x) 
\end{array}\right . \label{FIELDT} \end{equation}
It then follows that the arithmetic average
$\langle O(x) \rangle_{(m_q)}^{WA}=
\frac{1}{2}[\langle O(x) \rangle_{(r,m_q)}+
\langle O(x) \rangle_{(-r,m_q)}]$
is free of O($a$) discretization effects, because from the above 
$r$-parity considerations one gets
\beq
\langle O(x) \rangle_{(m_q)}^{WA}=\zeta^{O}_{O}(r)
\langle O(x) \rangle^{\rm{cont}}_{(m_q)}+{\rm{O}}(a^2)\, .
\label{WILAV}
\eeq
From this relation the O($a$) improvement of $WA$'s of hadronic masses and 
on-shell matrix elements can be immediately proved~\cite{FR}.

\section{TWISTED-MASS LATTICE QCD}

Wilson averaging can be straightforwardly extended to tm-LQCD~\cite{FR}.
To this end it is convenient to write the fermionic 
tm-LQCD  action for an $SU(2)_{\rm{f}}$ mass degenerate doublet in the form
\beqn
&&\hspace{-.5cm}S_{\rm{F}}^{\rm{L,tm}}=
a^4 \sum_x\,\bar\psi(x) 
[\frac{1}{2}\sum_\mu\gamma_\mu(\nabla^\star_\mu+ \nabla_\mu)+
\label{PHYSCHI}\\&&\hspace{-.5cm}
+(-r\frac{a}{2}\sum_\mu\nabla^\star_\mu\nabla_\mu+M_{\rm{cr}})
e^{-i\omega\gamma_5\tau_3}+m_q]\psi(x)\, ,
\nonumber\eeqn  
for in this quark basis (physical basis) the fermionic mass term is real.

Since the transformation~(\ref{SPUR}) is still a spurionic symmetry 
of~(\ref{PHYSCHI}) and it does not affect the twisting angle $\omega$, 
the whole line of arguments developed above goes through with 
correlators and derived quantities only having (at finite $a$) an extra 
dependence upon the label $\omega$.

While working with $\omega\neq 0$ solves all the problems related to 
spurious modes of the WD operator, it implies a breaking of the parity 
(and isospin) symmetry. However, if the parity operation, ${\cal P}$, is 
accompanied by a change of sign of $\omega$, the action~(\ref{PHYSCHI}) 
remains invariant. Consequently the spurionic symmetry 
${\cal P}\times (\omega\to-\omega)$ can be used to label states and operators
with a binary quantum number, which in the continuum limit (where by 
universality the $\omega$ dependence of lattice quantities drops out) 
is to be identified with the physical parity. For instance, the eigenstates of 
the lattice transfer matrix can always be taken to satisfy the formula
\beqn
{\cal{P}}\,|h,n,{\bf k}\>^{(\omega)}_{(r,m_q)}=
\eta_{h,n}|h,n,-{\bf k}\>^{(-\omega)}_{(r,m_q)}\, ,
\,\,\eta_{h,n}^2=1 \, ,\nonumber\eeqn
while the corresponding energy eigenvalues can be proved to obey the relations
\beq
E_{h,n}({\bf k};\omega,r,m_q) = 
E_{h,n}(\pm{\bf k};\pm\omega,r,m_q)\, .\label{EKO}
\eeq

\subsection{A special case: $\omega=\pm\pi/2$}
\label{sec:ASC}

The choice $\omega=\pm\pi/2$ in the tm-LQCD action~(\ref{PHYSCHI}) is 
especially worth mentioning, because all quantities that are even under 
$\omega\rightarrow -\omega$ are O($a$) improved with no need of any averaging.
This follows from the fact that for the particular value 
$\omega=\pm \pi/2$ a sign inversion of the twisting angle 
is equivalent (mod\,($2\pi)$) to a shift by $\pi$. This operation is in 
turn the same as inverting the sign of $r$: quantities 
even in $\omega$ are hence also even in $r$. As a result the two
terms entering the $WA$ are identical and
averaging is unnecessary to get O($a$) improvement. 
Automatically O($a$) improved quantities are e.g.~hadron masses and 
on-shell matrix elements at zero three-momentum. 
More examples are discussed in~\cite{FR}.

Another remarkable fact about the choice $\pm\pi/2$ is that it is possible to 
get an O($a$) improved estimate of $F_\pi$ which requires neither Wilson 
averaging nor the computation of any renormalization factor. To this end 
it is enough to use the 1-point split axial current ($x'\equiv x+a\hat{\mu}$)
\beqn
&&\hspace{-.5cm}\hat{A}_\mu^1(x)^{1-{\rm{pt}}}= 
\frac{1}{2}[\bar\psi(x)\frac{\tau_1}{2}
\gamma_\mu\gamma_5 U_\mu(x)\psi(x')+\nonumber\\
&&\hspace{-.5cm}+\bar\psi(x')\frac{\tau_1}{2}
\gamma_\mu\gamma_5 U^\dagger_\mu(x)\psi(x)]+
\nonumber\\
&&\hspace{-.5cm}-\frac{r}{2}[\bar\psi(x)\frac{\tau_2}{2}
U_\mu(x)\psi(x')-
\bar\psi(x')\frac{\tau_2}{2}
U^\dagger_\mu(x)\psi(x)]\, ,\nonumber\eeqn
which is exactly conserved at $m_q=0$, and observe that only zero 
three-momentum lattice correlators need be evaluated to extract $F_\pi$.

\section{CONCLUSIONS AND OUTLOOK}

The strategy we have proposed, besides ensuring O($a$) improvement, 
has the virtue of being very flexible and leaves the freedom 
of regularizing different flavours with Wilson terms of different chiral 
phases. One can prove~\cite{FRTWO} that,
without loosing O($a$) improvement, this freedom can be 
exploited to have a real and positive fermionic determinant, while at the 
same time canceling all finite and infinite contributions due to mixing with 
operators of ``wrong chirality'' (i.e.~those due to the breaking of 
chiral symmetry induced by the presence of the Wilson term in the 
lattice action) in the calculation of the CP-conserving matrix elements 
of the effective weak Hamiltonian. In particular no power divergent mixings
survive in the amplitudes relevant for the $\Delta I =1/2$-rule. 
\vspace{.2cm}

{\bf Acknowledgements} - I wish to thank the LOC of LAT2003 for the 
impeccable organization and lively atmosphere of the meeting.

\end{document}